%% file: pecaut.tex
\documentclass{iau}
\usepackage{graphicx}
\usepackage{natbib}
\usepackage{subfig}

\input{shortcuts.tex}

\title[Young Stellar Colors and Spectral Types] 
{Anomalous Spectral Types and Intrinsic Colors of Young Stars}

\author[Pecaut]   
{Mark J. Pecaut}

\affiliation{Rockhurst University, 1100 Rockhurst Rd., Kansas City, MO 64110\\email: {\tt mark.pecaut@rockhurst.edu}}

\pubyear{2015}
\volume{314}  
\pagerange{119--126}
\jname{Young Stars \& Planets Near the Sun}
\editors{J. H. Kastner, B. Stelzer, \& S. A. Metchev, eds.}
\begin{document}

\maketitle

\begin{abstract}
We highlight differences in spectral types and intrinsic colors observed in
pre-main sequence (pre-MS) stars.  Spectral types of pre-MS stars are 
wavelength-dependent, with near-infrared spectra being 3-5 spectral 
sub-classes later than the spectral types determined from optical
spectra.  In addition, the intrinsic colors of young stars differ from that of
main-sequence stars at a given spectral type. 
We caution observers to adopt optical spectral types over near-infrared
types, since Hertzsprung-Russell (H-R) diagram positions derived from 
optical spectral types provide consistency between dynamical masses and 
theoretical evolutionary tracks.  We also urge observers to 
deredden pre-MS stars with tabulations of intrinsic colors specifically 
constructed for young stars, since their unreddened colors differ from 
that of main sequence dwarfs.  Otherwise, $V$-band extinctions as 
much as $\sim$0.6 mag erroneously higher than the true extinction
may result, which would introduce systematic errors in the H-R diagram 
positions and thus bias the inferred ages.
\end{abstract}

\firstsection 
\section{Introduction}

Two of the most fundamental parameters of a star -- the effective temperature (\teff)
and luminosity -- are based on simple, easy-to understand data such as the spectral type,
extinction, and bolometric corrections.  Determining these should be relatively 
error-free, right?  Experience has taught observers that special care must be taken
when characterizing pre-main sequence (pre-MS) stars, as these young stars require more attention
than that of main-sequence dwarfs.

When observers desire to characterize a star's properties, they normally start with 
the spectral type.  The spectral type of the star is determined by comparing 
characteristics of the spectrum with spectral standard stars.  In doing this
we can obtain an estimate of the temperature and a gross estimate of the 
surface gravity of the star.  To quantify the extinction and reddening, the 
observer will compare the target star's observed colors to tabulated intrinsic 
colors of stars of the same spectral type to determine a color excess and use 
a total-to-selective extinction ratio value to estimate the extinction.  Once the 
extinction has been quantified, a distance can be estimated (if not known), by 
assuming an age and consulting a theoretical isochrone, or assuming the star is on 
the main-sequence and calculating a main-sequence distance.  Alternatively, if a 
trigonometric or kinematic parallax is known, we may compute the luminosity of the 
star.  Normally, the point of these calculations is to place the star on the 
Hertzsprung-Russell (H-R) diagram and compare it to theoretical evolutionary models to 
estimate an age and mass.

Though this process seems fairly straightforward, there are many assumptions and 
systematic effects which can creep in and result in systematic errors in the fundamental 
parameters, such as temperature and luminosity (which are relatively model-free), and 
therefore the parameters derived from the evolutionary models.  Assuming we are able to 
determine the spectral type with perfect fidelity, there are uncertainties in the tables 
which relate the spectral type, \teff, intrinsic color and bolometric correction.  There 
are also many different varieties of such tables, with slight variations in their 
intrinsic colors, underlying temperature scale, and bolometric corrections.  Some tables 
even contain self-inconsistent values for the bolometric corrections and the bolometric 
luminosity of the Sun \citep{torres2010}\footnote{There are even a variety of
values used for the bolometric magnitude of the Sun.  Here we adopt 
$M_{bol,\odot}=4.7554\pm0.0004$, as advocated in \citet{mamajek2012b}.}!

For populations of young stars this presents many problems because systematic errors 
in the fundamental properties of young, pre-MS stars will propagate into 
systematic errors in masses and ages (see \citealt{soderblom2014} for a full discussion
of ages of young stars).  This can skew the inferred evolutionary 
lifetime of gas-rich disks.   Since gas giant planets 
can only form when gas is present in the circumstellar disk, these ages are also used 
to constrain giant planet formation timescales.  Problems are also present when 
individual stars are mischaracterized.  If a young star hosts a directly-imaged 
substellar object, the mass of the substellar object is estimated by comparing 
the luminosity and assumed age of the object with evolutionary models.  Systematic 
effects in assumed ages can then propagate to wrong assumptions about the model-derived 
masses (e.g., $\kappa$~And~b; \citealt{hinkley2013,bonnefoy2014}), which may misdirect 
planet formation theories.  Thus, systematic errors in individual parameters, such as 
spectral types and intrinsic colors, can propagate down and fundamentally limit our 
ability to test star and planet formation theories.

\section{Spectral Types}

The spectral type of a target is one of the most useful measurements, since many other 
stellar properties are usually derived with some dependence on the spectral type.  Young 
stars, like most stars, are typed by comparing their spectra with that of spectral 
standards, and this has historically been performed using optical spectra.  However, many 
low-mass stars are brighter in the near-infrared (NIR) and so it seems completely 
reasonable to perform this same measurement with NIR spectra as well.  Two 
interesting cases are that of TW~Hya and V4046~Sgr.

TW~Hya, one of the most well-studied classical T-Tauri stars and a member of the 
youngest nearby moving group that bears its name (the TW Hydra Association),
has typically been assigned a temperature type of K7 using optical spectra 
(K8IVe, \citealt{pecaut2013}; K6Ve, \citealt{torres2006}; K6e, \citealt{hoff1998}; 
K7e, \citealt{delareza1989}; K7 Ve, \citealt{herbig1978}).  However, 
\citet{vacca2011} assigned a type of M2.5V using NIR spectra from SpeX, which implied a 
very young age of $\sim$3~Myr instead of the much older, more often quoted age of 
$\sim$~10 Myr \citep{barrado2006}.  This discrepancy is a source of great confusion 
-- which spectral type should one adopt when characterizing young stars?

V4046~Sgr, a young binary member of the $\beta$~Pictoris moving group harboring a 
gas-rich disk of its own, has also been typed in both the optical and NIR and the same 
effect is observed - the NIR spectral type is about 3-5 subtypes later than the optical 
spectral type \citep{kastner2014}.  However, unlike TW~Hya, V4046~Sgr has dynamical 
mass constraints from radial velocities and gas dynamics \citep{rosenfeld2012}.  
\citet{kastner2014} have placed these two young stars on the H-R diagram assuming 
the optical spectral types in one case and the NIR spectral types in another case.  
Comparing these two sets of H-R diagram positions with theoretical evolutionary models, 
the NIR spectral types are inconsistent with the dynamical mass constraints and 
\citet{kastner2014} thus urged caution against using the NIR spectral types on young stars.

\citet{stauffer2003} have studied this wavelength-dependent spectral type effect among 
the zero-age main sequence K dwarfs in the Pleiades \citep[$\sim$135~Myr;][]{bell2014}. 
The \citet{stauffer2003} study found that the spectral 
type of Pleiades K-type stars were systematically $\sim$1 subtype later in the red 
optical spectra than the blue optical spectra.  Furthermore, they did not observe 
this spectral type anomaly in members of the older Praesepe cluster 
\citep[$\sim$650-800~Myr;][]{gaspar2009,bell2014,brandt2015}.  \citet{stauffer2003} 
argued that spots were a major factor in this effect, and concluded that there 
must be more than one photospheric temperature present in the Pleiades K-dwarfs.  
Pre-main sequence stars are magnetically very active as well, and observations indicate 
large filling factors on their surfaces \citep{berdyugina2005}, consistent with this 
effect.

\section{Intrinsic Colors}

Many studies in the past two decades have pointed out that stellar intrinsic colors of 
young stars are different than that of typical main sequence dwarfs.  
\citet{gullbring1998} had noted this for both classical and weak-lined T-Tauri 
stars in Taurus.  \citet{dario2010} have noted this in the Orion Nebula Cluster (ONC) and
dereddened the very young ONC cluster members using a specially constructed spectral-type 
color sequence for young stars.  However, their spectral type-color sequence was not 
published as part of their study.  \citet{luhman1999b} and \citet{luhman2010,luhman2010e} 
went much further and constructed a color-spectral type sequence for pre-MS
late K- and M-type stars and brown dwarfs.  Most recently, the \citet{pecaut2013} study 
released a comprehensive tabulation of the intrinsic colors for pre-MS
stars from F-type to late M-type with the most popular photometric bands -- 
Johnson--Cousins $BVI_C$, 2MASS $JHK_S$ and the recently available WISE $W1$, $W2$, 
$W3$ and $W4$ bands.  In addition, we fit the observed spectral energy 
distributions to Phoenix--NextGen synthetic spectra \citep{allard2012} to infer an 
effective temperature (\teff) and bolometric correction (BC) scale for pre-MS
stars.  This young spectral type--color--\teff--BC scale was constructed using 
all the known (as of July 2013) members of the nearest moving groups -- the 
$\eta$~Cha Cluster, the TW Hydra Association (TWA), the $\beta$~Pictoris moving group and 
the Tucana--Horologium (Tuc-Hor) moving group.

\begin{figure}[ht]
\begin{center}
\subfloat{\includegraphics[width=0.50\columnwidth]{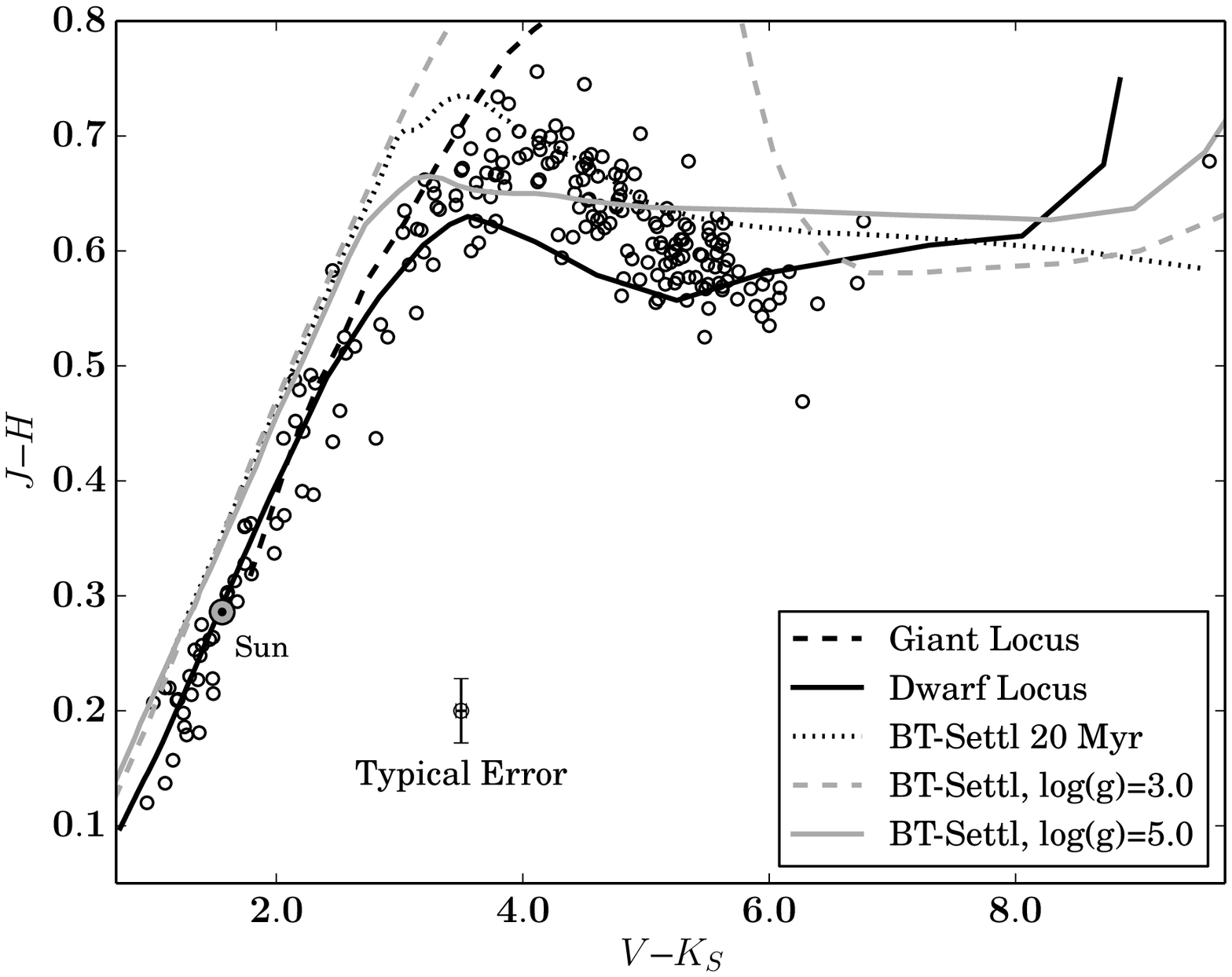}}
\subfloat{\includegraphics[width=0.50\columnwidth]{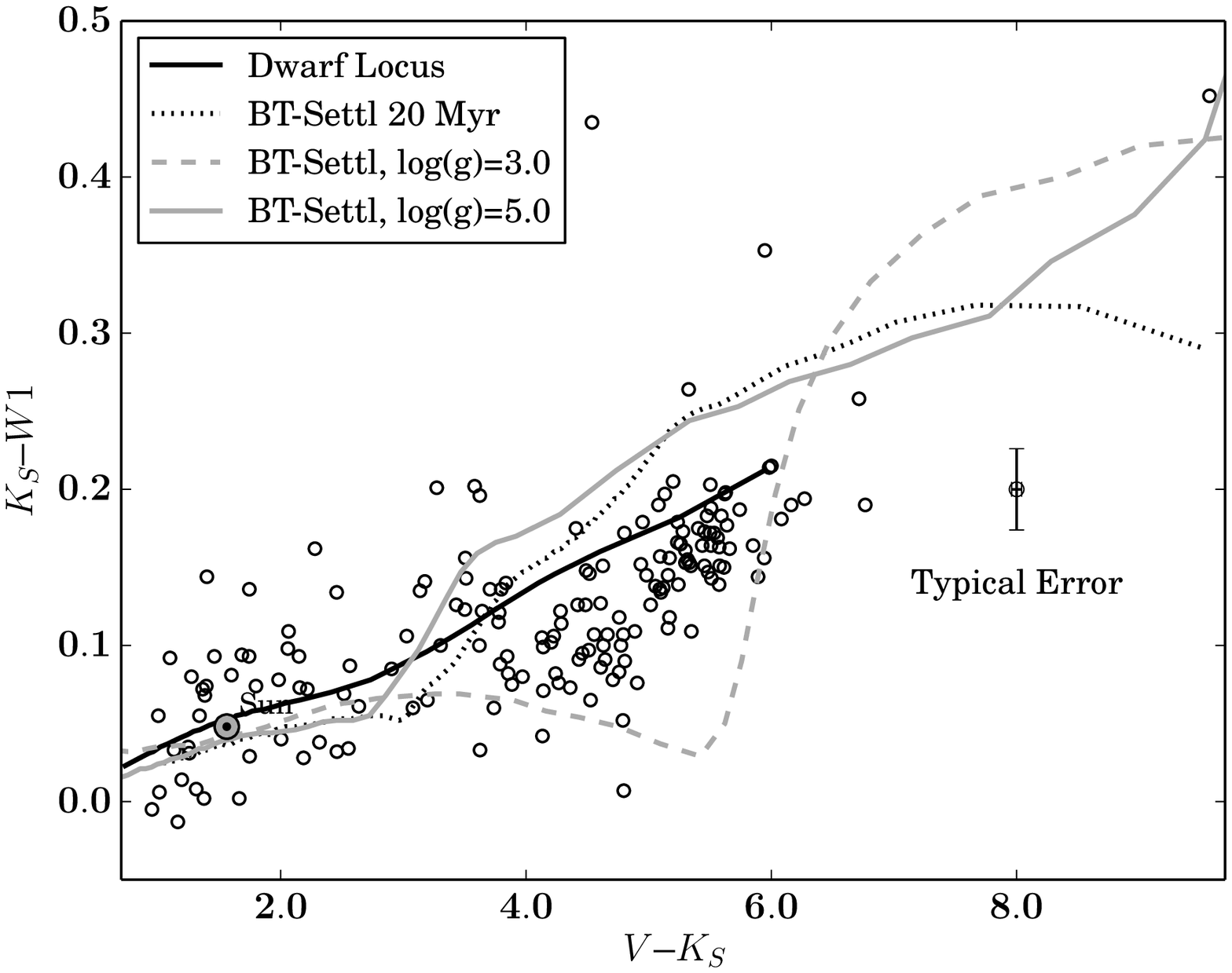}}
\caption{$V$--$K_S$ versus $J$--$H$ (left) and $V$--$K_S$ versus $K_S$--$W1$ (right) 
for young stars in the $\eta$~Cha cluster, the TW Hydra Association, the $\beta$~Pictoris 
moving group and the Tucana--Horologium moving group from the sample in \citet{pecaut2013},
with the addition of 129 new Tuc-Hor members from \citet{kraus2014}.  
The dwarf locus is adopted from \citet{pecaut2013} and the giant locus is 
adopted from \citet{bessell1998}.  20~Myr isochronal colors are constructed by adopting 
surface gravities from a \citet{baraffe1998} 20~Myr isochrone with synthetic colors from 
the BT-Settl models of \citet{allard2012}.}
\label{fig:color-color}
\end{center}
\end{figure}

Two color-color plots for young stars are shown in Figure~\ref{fig:color-color}.  We 
plot $V$--$K_S$ on the horizontal axis as a proxy for \teff\, against $J$--$H$ (left) 
and $K_S$--$W1$ (right).  The stars are predominantly clustered around the dwarf and 
giant sequence until the two diverge around $V$--$K_S$ of $\sim$4, where they lie 
between the dwarf and giant locus.  This strongly suggests surface gravity is a major 
factor in the deviation of intrinsic colors from that of dwarfs, since we expect that 
pre-MS stars will have surface gravities somewhere between that of dwarfs and 
giants.  

\begin{figure}[ht]
\begin{center}
\subfloat{\includegraphics[width=0.50\textwidth]{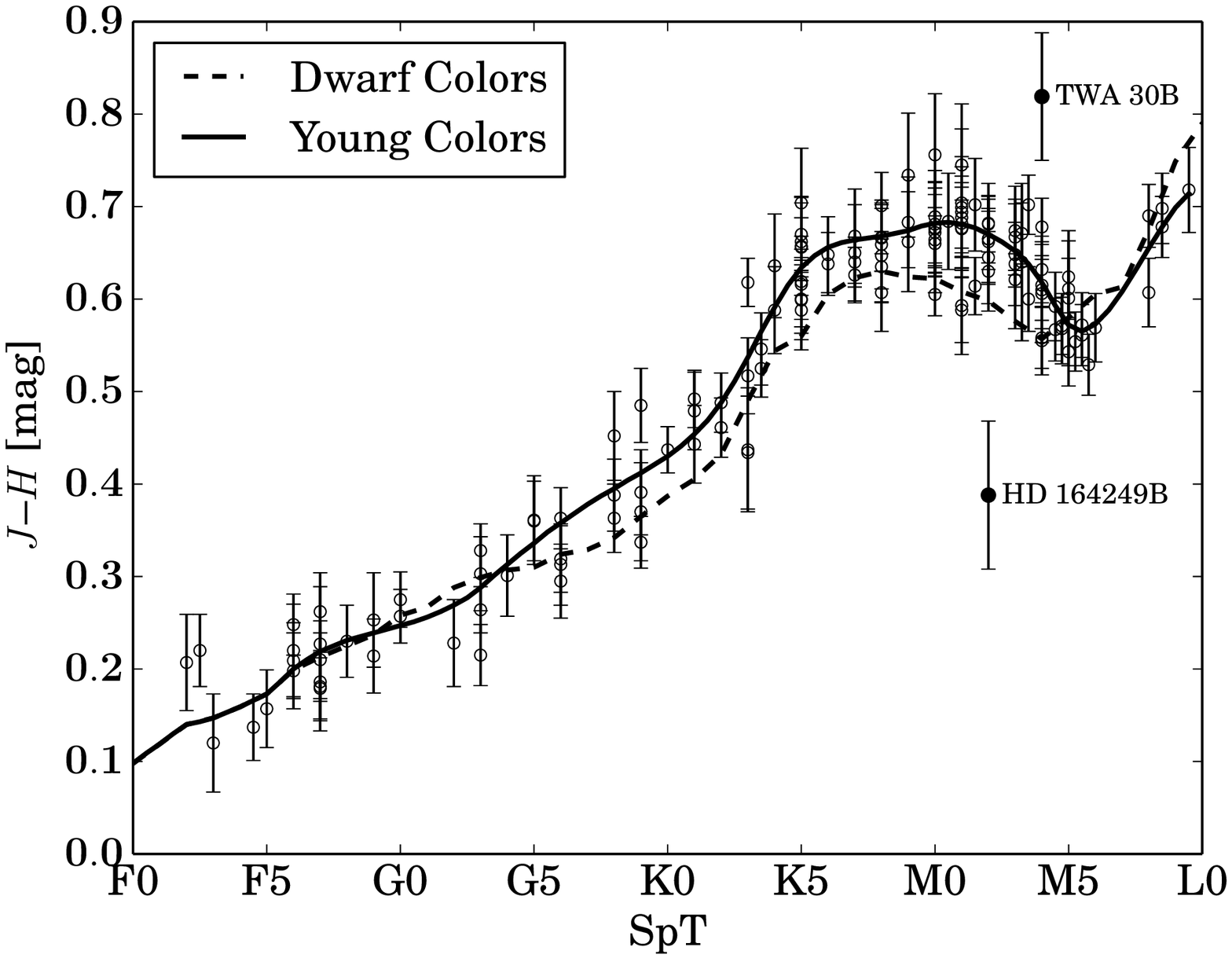}\label{fig:spt_jh} }
\subfloat{\includegraphics[width=0.50\textwidth]{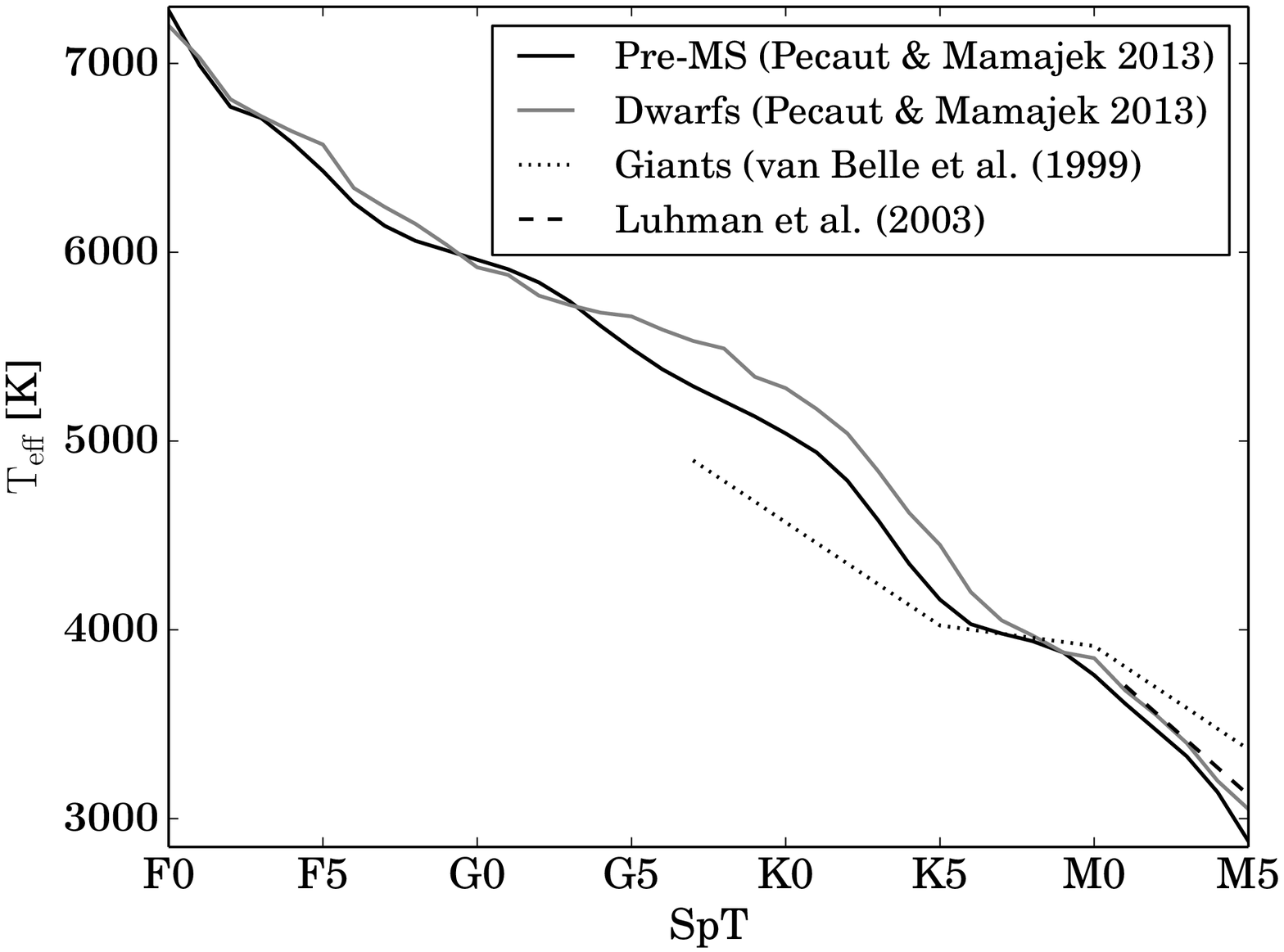}\label{fig:spt_teff} }
\caption{Left: Spectral type versus $J$--$H$ for young stars in the $\eta$~Cha cluster, 
the TW Hydra Association, the $\beta$~Pictoris moving group and the Tucana--Horologium 
moving group from the sample in \citet{pecaut2013}, with the addition of 129 new 
Tuc-Hor members from \citet{kraus2014}.
Right: Spectral type versus effective temperature for pre-MS stars 
\citep{pecaut2013}, dwarf stars \citep{pecaut2013}, giants \citep{vanbelle1999},
and the M-type \teff\, scale from \citet{luhman2003}.  The pre-MS
temperature scale shown is $\sim$200K cooler than the dwarf scale for types
$\sim$G5--K5.}
\end{center}
\end{figure}

An indication of the importance of accounting for the difference between pre-MS
colors and dwarf colors is shown in Figure~\ref{fig:spt_jh}.  The $J$--$H$
colors of young stars are significantly redder than the dwarf colors at a given spectral type.
If $J$--$H$ dwarf colors were used to estimate extinction for an unreddened M0 star, 
it would erroneously appear to have an $A_V \simeq 0.6$~mag of extinction!  If dwarf 
colors are used to deredden a population of young stars, a systematic bias may be 
introduced, depending on the particular color used, which may cause their luminosities 
to be systematically overestimated and thus their ages would be systematically 
underestimated.

A careful look at the temperature scale of pre-MS stars is warranted, since a star's 
adopted spectral type is used to infer the \teff.  A frequently used
temperature scale for M-type stars is the scale of \citet{luhman2003}, which is intermediate
between dwarfs and giants.  However, in the past few years two important developments have 
been made available: (1) with the release of the {\it 2MASS} and {\it WISE} catalogs 
\citep{skrutskie2006,cutri2012a}, photometry is available which covers large
sections of the star's spectral energy distributions (SED) and (2) high-quality synthetic
spectra are available from the Phoenix-NextGen group \citep{allard2012} which include 
updated molecular opacities and model low-temperature stars and brown dwarfs more
successfully than ever possible previously.  Thus it is possible to obtain
tight constraints on the \teff\, of an individual unreddened pre-MS member
of one of the nearby, young moving groups through fitting the observed SED with synthetic
models.  \citet{pecaut2013} fit Phoenix-NextGen BT-Settl models of 
\citet{allard2012} to the observed SEDs of the members of $\eta$~Cha, TWA, 
$\beta$~Pic and Tuc-Hor to tie their updated spectral type intrinsic color tabulation to a 
\teff-BC system.    The results of this \teff\, scale are shown in Figure~\ref{fig:spt_teff}.  
Pre-MS stars are $\sim$200K systematically cooler than their main 
sequence counterparts at a given spectral type, for spectral types $\sim$G5--K5.  

\section{Conclusions}
When characterizing the basic observations and properties of young stars, such as spectral 
type, \teff, reddening and extinction, care must be exercised to avoid systematic errors
and biases.  It is important that optical spectra be used to constrain the spectral type
of the star, since these seem to best represent the effective temperature of the stars. 
Some evidence points to spots as a factor in wavelength-dependent spectral types, discussed
in detail by \citet{gullbring1998} and \citet{stauffer2003}.

When considering or using tabulations of intrinsic colors, it is important to adopt 
intrinsic colors for pre-MS stars when the stars are still contracting
to the main sequence.  Comparisons with dwarf and giant colors as well as theoretical
synthetic spectra of different surface gravity points to surface gravity as an important
effect altering the intrinsic colors of pre-MS stars.  
If individual extinctions are mis-estimated using dwarf colors, stars placed on the H-R 
diagram may have their luminosities systematically over-estimated and the 
population may appear younger simply because observers are unable to account for 
reddening and extinction properly.  This clearly inhibits our ability to 
accurately test theoretical evolutionary models and may systematically bias ages, 
which has consequences reaching down to mis-estimating age spreads in star-forming 
regions, underestimating the evolutionary timescales of disks and planet formation 
timescales, and systematic errors when inferring the initial mass function (IMF) in 
pre-MS stellar populations.

\bibliographystyle{apj}
\bibliography{pecaut}

\end{document}

%% file: shortcuts.tex
\newcommand{\simless}{\mathbin{\lower 3pt\hbox
     {$\rlap{\raise 5pt\hbox{$\char'074$}}\mathchar"7218$}}}
\newcommand{\simgreat}{\mathbin{\lower 3pt\hbox
     {$\rlap{\raise 5pt\hbox{$\char'076$}}\mathchar"7218$}}}

\newcommand{\teff}{T$_{\rm eff}$}

%% file: pecaut.bbl
\begin{thebibliography}{32}
\expandafter\ifx\csname natexlab\endcsname\relax\def\natexlab#1{#1}\fi

\bibitem[{{Allard} {et~al.}(2012){Allard}, {Homeier}, \&
  {Freytag}}]{allard2012}
{Allard}, F., {Homeier}, D., \& {Freytag}, B. 2012, Royal Society of London
  Philosophical Transactions Series A, 370, 2765

\bibitem[{{Baraffe} {et~al.}(1998){Baraffe}, {Chabrier}, {Allard}, \&
  {Hauschildt}}]{baraffe1998}
{Baraffe}, I., {Chabrier}, G., {Allard}, F., \& {Hauschildt}, P.~H. 1998, \aap,
  337, 403

\bibitem[{{Barrado Y Navascu{\'e}s}(2006)}]{barrado2006}
{Barrado Y Navascu{\'e}s}, D. 2006, \aap, 459, 511

\bibitem[{{Bell} {et~al.}(2014){Bell}, {Rees}, {Naylor}, {Mayne}, {Jeffries},
  {Mamajek}, \& {Rowe}}]{bell2014}
{Bell}, C.~P.~M., {Rees}, J.~M., {Naylor}, T., {Mayne}, N.~J., {Jeffries},
  R.~D., {Mamajek}, E.~E., \& {Rowe}, J. 2014, \mnras, 445, 3496

\bibitem[{{Berdyugina}(2005)}]{berdyugina2005}
{Berdyugina}, S.~V. 2005, Living Reviews in Solar Physics, 2, 8

\bibitem[{{Bessell} {et~al.}(1998){Bessell}, {Castelli}, \&
  {Plez}}]{bessell1998}
{Bessell}, M.~S., {Castelli}, F., \& {Plez}, B. 1998, \aap, 333, 231

\bibitem[{{Bonnefoy} {et~al.}(2014){Bonnefoy}, {Currie}, {Marleau},
  {Schlieder}, {Wisniewski}, {Carson}, {Covey}, {Henning}, {Biller}, {Hinz},
  {Klahr}, {Marsh Boyer}, {Zimmerman}, {Janson}, {McElwain}, {Mordasini},
  {Skemer}, {Bailey}, {Defr{\`e}re}, {Thalmann}, {Skrutskie}, {Allard},
  {Homeier}, {Tamura}, {Feldt}, {Cumming}, {Grady}, {Brandner}, {Helling},
  {Witte}, {Hauschildt}, {Kandori}, {Kuzuhara}, {Fukagawa}, {Kwon}, {Kudo},
  {Hashimoto}, {Kusakabe}, {Abe}, {Brandt}, {Egner}, {Guyon}, {Hayano},
  {Hayashi}, {Hayashi}, {Hodapp}, {Ishii}, {Iye}, {Knapp}, {Matsuo}, {Mede},
  {Miyama}, {Morino}, {Moro-Martin}, {Nishimura}, {Pyo}, {Serabyn}, {Suenaga},
  {Suto}, {Suzuki}, {Takahashi}, {Takami}, {Takato}, {Terada}, {Tomono},
  {Turner}, {Watanabe}, {Yamada}, {Takami}, \& {Usuda}}]{bonnefoy2014}
{Bonnefoy}, M., {et~al.} 2014, \aap, 562, A111

\bibitem[{{Brandt} \& {Huang}(2015)}]{brandt2015}
{Brandt}, T.~D., \& {Huang}, C.~X. 2015, ArXiv e-prints

\bibitem[{{Cutri} {et~al.}(2012){Cutri}, {Skrutskie}, {van Dyk}, {Beichman},
  {Carpenter}, {Chester}, {Cambresy}, {Evans}, {Fowler}, {Gizis}, {Howard},
  {Huchra}, {Jarrett}, {Kopan}, {Kirkpatrick}, {Light}, {Marsh}, {McCallon},
  {Schneider}, {Stiening}, {Sykes}, {Weinberg}, {Wheaton}, {Wheelock}, \&
  {Zacharias}}]{cutri2012a}
{Cutri}, R.~M., {et~al.} 2012, VizieR Online Data Catalog, 2281, 0

\bibitem[{{Da Rio} {et~al.}(2010){Da Rio}, {Robberto}, {Soderblom}, {Panagia},
  {Hillenbrand}, {Palla}, \& {Stassun}}]{dario2010}
{Da Rio}, N., {Robberto}, M., {Soderblom}, D.~R., {Panagia}, N., {Hillenbrand},
  L.~A., {Palla}, F., \& {Stassun}, K.~G. 2010, \apj, 722, 1092

\bibitem[{{de la Reza} {et~al.}(1989){de la Reza}, {Torres}, {Quast},
  {Castilho}, \& {Vieira}}]{delareza1989}
{de la Reza}, R., {Torres}, C.~A.~O., {Quast}, G., {Castilho}, B.~V., \&
  {Vieira}, G.~L. 1989, \apjl, 343, L61

\bibitem[{{G{\'a}sp{\'a}r} {et~al.}(2009){G{\'a}sp{\'a}r}, {Rieke}, {Su},
  {Balog}, {Trilling}, {Muzzerole}, {Apai}, \& {Kelly}}]{gaspar2009}
{G{\'a}sp{\'a}r}, A., {Rieke}, G.~H., {Su}, K.~Y.~L., {Balog}, Z., {Trilling},
  D., {Muzzerole}, J., {Apai}, D., \& {Kelly}, B.~C. 2009, \apj, 697, 1578

\bibitem[{{Gullbring} {et~al.}(1998){Gullbring}, {Hartmann}, {Briceno}, \&
  {Calvet}}]{gullbring1998}
{Gullbring}, E., {Hartmann}, L., {Briceno}, C., \& {Calvet}, N. 1998, \apj,
  492, 323

\bibitem[{{Herbig}(1978)}]{herbig1978}
{Herbig}, G.~H. 1978, {Can Post-T Tauri Stars Be Found?}, ed. L.~V. {Mirzoyan},
  171

\bibitem[{{Hinkley} {et~al.}(2013){Hinkley}, {Pueyo}, {Faherty}, {Oppenheimer},
  {Mamajek}, {Kraus}, {Rice}, {Ireland}, {David}, {Hillenbrand}, {Vasisht},
  {Cady}, {Brenner}, {Veicht}, {Nilsson}, {Zimmerman}, {Parry}, {Beichman},
  {Dekany}, {Roberts}, {Roberts}, {Baranec}, {Crepp}, {Burruss}, {Wallace},
  {King}, {Zhai}, {Lockhart}, {Shao}, {Soummer}, {Sivaramakrishnan}, \&
  {Wilson}}]{hinkley2013}
{Hinkley}, S., {et~al.} 2013, \apj, 779, 153

\bibitem[{{Hoff} {et~al.}(1998){Hoff}, {Henning}, \& {Pfau}}]{hoff1998}
{Hoff}, W., {Henning}, T., \& {Pfau}, W. 1998, \aap, 336, 242

\bibitem[{{Kastner} {et~al.}(2015){Kastner}, {Rapson}, {Sargent}, {Smith}, \&
  {Rayner}}]{kastner2014}
{Kastner}, J.~H., {Rapson}, V., {Sargent}, B., {Smith}, C.~T., \& {Rayner}, J.
  2015, in Cambridge Workshop on Cool Stars, Stellar Systems, and the Sun,
  Vol.~18, Cambridge Workshop on Cool Stars, Stellar Systems, and the Sun, ed.
  G.~T. {van Belle} \& H.~C. {Harris}, 313--320

\bibitem[{{Kraus} {et~al.}(2014){Kraus}, {Shkolnik}, {Allers}, \&
  {Liu}}]{kraus2014}
{Kraus}, A.~L., {Shkolnik}, E.~L., {Allers}, K.~N., \& {Liu}, M.~C. 2014, \aj,
  147, 146

\bibitem[{{Luhman}(1999)}]{luhman1999b}
{Luhman}, K.~L. 1999, \apj, 525, 466

\bibitem[{{Luhman} {et~al.}(2010{\natexlab{a}}){Luhman}, {Allen}, {Espaillat},
  {Hartmann}, \& {Calvet}}]{luhman2010e}
{Luhman}, K.~L., {Allen}, P.~R., {Espaillat}, C., {Hartmann}, L., \& {Calvet},
  N. 2010{\natexlab{a}}, \apjs, 189, 353

\bibitem[{{Luhman} {et~al.}(2010{\natexlab{b}}){Luhman}, {Allen}, {Espaillat},
  {Hartmann}, \& {Calvet}}]{luhman2010}
---. 2010{\natexlab{b}}, \apjs, 186, 111

\bibitem[{{Luhman} {et~al.}(2003){Luhman}, {Stauffer}, {Muench}, {Rieke},
  {Lada}, {Bouvier}, \& {Lada}}]{luhman2003}
{Luhman}, K.~L., {Stauffer}, J.~R., {Muench}, A.~A., {Rieke}, G.~H., {Lada},
  E.~A., {Bouvier}, J., \& {Lada}, C.~J. 2003, \apj, 593, 1093

\bibitem[{{Mamajek}(2012)}]{mamajek2012b}
{Mamajek}, E.~E. 2012, \apjl, 754, L20

\bibitem[{{Pecaut} \& {Mamajek}(2013)}]{pecaut2013}
{Pecaut}, M.~J., \& {Mamajek}, E.~E. 2013, \apjs, 208, 9

\bibitem[{{Rosenfeld} {et~al.}(2012){Rosenfeld}, {Andrews}, {Wilner}, \&
  {Stempels}}]{rosenfeld2012}
{Rosenfeld}, K.~A., {Andrews}, S.~M., {Wilner}, D.~J., \& {Stempels}, H.~C.
  2012, \apj, 759, 119

\bibitem[{{Skrutskie} {et~al.}(2006){Skrutskie}, {Cutri}, {Stiening},
  {Weinberg}, {Schneider}, {Carpenter}, {Beichman}, {Capps}, {Chester},
  {Elias}, {Huchra}, {Liebert}, {Lonsdale}, {Monet}, {Price}, {Seitzer},
  {Jarrett}, {Kirkpatrick}, {Gizis}, {Howard}, {Evans}, {Fowler}, {Fullmer},
  {Hurt}, {Light}, {Kopan}, {Marsh}, {McCallon}, {Tam}, {Van Dyk}, \&
  {Wheelock}}]{skrutskie2006}
{Skrutskie}, M.~F., {et~al.} 2006, \aj, 131, 1163

\bibitem[{{Soderblom} {et~al.}(2014){Soderblom}, {Hillenbrand}, {Jeffries},
  {Mamajek}, \& {Naylor}}]{soderblom2014}
{Soderblom}, D.~R., {Hillenbrand}, L.~A., {Jeffries}, R.~D., {Mamajek}, E.~E.,
  \& {Naylor}, T. 2014, Protostars and Planets VI, 219

\bibitem[{{Stauffer} {et~al.}(2003){Stauffer}, {Jones}, {Backman}, {Hartmann},
  {Barrado y Navascu{\'e}s}, {Pinsonneault}, {Terndrup}, \&
  {Muench}}]{stauffer2003}
{Stauffer}, J.~R., {Jones}, B.~F., {Backman}, D., {Hartmann}, L.~W., {Barrado y
  Navascu{\'e}s}, D., {Pinsonneault}, M.~H., {Terndrup}, D.~M., \& {Muench},
  A.~A. 2003, \aj, 126, 833

\bibitem[{{Torres} {et~al.}(2006){Torres}, {Quast}, {da Silva}, {de La Reza},
  {Melo}, \& {Sterzik}}]{torres2006}
{Torres}, C.~A.~O., {Quast}, G.~R., {da Silva}, L., {de La Reza}, R., {Melo},
  C.~H.~F., \& {Sterzik}, M. 2006, \aap, 460, 695

\bibitem[{{Torres}(2010)}]{torres2010}
{Torres}, G. 2010, \aj, 140, 1158

\bibitem[{{Vacca} \& {Sandell}(2011)}]{vacca2011}
{Vacca}, W.~D., \& {Sandell}, G. 2011, \apj, 732, 8

\bibitem[{{van Belle} {et~al.}(1999){van Belle}, {Lane}, {Thompson}, {Boden},
  {Colavita}, {Dumont}, {Mobley}, {Palmer}, {Shao}, {Vasisht}, {Wallace},
  {Creech-Eakman}, {Koresko}, {Kulkarni}, {Pan}, \& {Gubler}}]{vanbelle1999}
{van Belle}, G.~T., {et~al.} 1999, \aj, 117, 521

\end{thebibliography}
